\documentclass{mn2e}
\usepackage{psfig}

\newcommand\lmc{LMC~X-3}
\newcommand\etal{et al. }
\newcommand\ie{i.e. }

\newcommand\xmm{{\it XMM-Newton}}
\newcommand\chandra{{\it Chandra}}

\newcommand\nh{\rm N_{H}}

\begin{document}

   \title{\xmm\ RGS spectroscopy of 
\lmc}
   \author[Page, et al.]{M.J. Page$^{1}$,
           R. Soria$^{1}$,
           K. Wu$^{1}$,
           K.O. Mason$^{1}$,
           F.A. Cordova$^{2}$ and
           W.C. Priedhorsky$^{3}$\\
$^{1}$Mullard Space Science Laboratory, University College London,
Holmbury St Mary, Dorking, Surrey, RH5 6NT, UK\\
$^{2}$Department of Physics, University of California, Riverside, CA93106, 
USA\\
$^{3}$Los Alamos National Laboratory, Los Alamos, NM 87545, USA\\
          }

\maketitle
\begin{abstract}
We present soft X-ray spectroscopy of the black hole binary \lmc\ from the
\xmm\ Reflection Grating Spectrometer. The observations span the full range of
spectral states seen in \lmc. The spectra are completely dominated by continuum
emission, and the neutral absorbing column measured from the O~I edge ($\nh =
3.8^{+0.8}_{-0.7} \times 10^{20}$~cm$^{-2}$) is consistent with the Galactic
interstellar column density towards \lmc. We find no evidence for variability
of the neutral absorbing column. 
We also constrain the ionized column density
using the upper limits to the equivalent width of the O~II -- O~VIII K-shell
resonance lines: we find that the equivalent hydrogen column density of gas
in which O is partially ionized is $< 8\times 10^{20}$~cm$^{-2}$. 
From this upper limit we can rule out a
line driven stellar wind as the power source for the X-ray emission of \lmc\
except when it is faint.
At wavelengths longward of the peak
emission the spectral shape is well described by a multi-temperature
disk-blackbody spectrum; the powerlaw component which dominates at shorter
wavelengths does not continue longward of the disk-blackbody peak. This implies
that the multi-temperature disk-blackbody component supplies the seed photons
which are Compton upscattered in the hot corona, consistent with the standard
paradigm for black hole X-ray binary spectral states.
\end{abstract}

%

\maketitle

\section{Introduction}
\label{sec:introduction}

The X-ray properties of accreting black hole candidates (BHCs) lead naturally 
to classification into
different spectral states \nocite{vanderklis94} (e.g. Van der Klis 1994). 
In their ``high/soft''
state, their X-ray spectra are dominated below 10 keV by a thermal
blackbody-like component with temperature $\sim 1$ keV. In their 
``low/hard'' state, this thermal component is absent, and
BHCs have powerlaw X-ray spectra. 
An ``intermediate'' state is also seen, in which
the thermal component is weaker. 
Individual BHCs can exhibit
all spectral states, but the fraction of time spent in the different
states varies considerably between objects: for example LMC X-1 has
only been observed in the high/soft state, while Cyg X-1 spends the
majority of its time in the low/hard state. 

Of all the persistent BHCs known, 
LMC X-3 is viewed through the smallest Galactic 
column density.
Until recently \lmc\ was thought to be permanently in a high/soft
state, but  
observations with the {\it Rossi X-ray Timing Explorer} ({\it RXTE}) 
have shown that \lmc\ changes to a low/hard state during periods
of low count rate \nocite{wilms01} (Wilms \etal 2001). The existence of state
transitions in \lmc\ has now been confirmed by further {\it RXTE}
observations \nocite{boyd00} (Boyd \etal 2000) and \xmm\ observations
\nocite{wu01} (Wu \etal 2001).  Wilms \etal (2001) \nocite{wilms01} argue 
that the long term 
variability and
consequent state changes could be due to an accretion disk 
wind-driven limit cycle 
(Shields \etal \nocite{shields86} 1986).

Because of the small column density in the direction of \lmc, it is
the best persistent BHC in which to study the soft X-ray spectrum at a
variety of spectral states.  The soft X-ray spectral region is
particularly important for understanding the accretion and mass transfer 
processes,
because winds from the accretion disk, or from the companion star will
imprint the soft X-ray spectrum with emission and absorption lines
and edges.  Furthermore, the standard paradigm for the thermal
component seen in the high and intermediate states of BHCs (namely that it is
the multi-temperature blackbody spectrum of an optically thick,
gemoetrically thin, accretion disk) has been poorly tested at soft
X-ray energies. This is because almost all instruments on the
previous generation of X-ray observatories (\ie before the launches of
\chandra\ and \xmm) had poor spectral resolution below 2 keV, leading to 
degeneracies between different spectral shapes and different amounts of 
absorption in spectral modelling. 

In this paper we present soft X-ray 
spectra of LMC X-3 from the  \xmm\ Reflection Grating Spectrometer (RGS). 
The spectra, which have a resolution of $\sim$70~m\AA, 
cover the full range of observed spectral states in this source. 

\section{Observations and data reduction}
\label{sec:observation}

\begin{figure}
\begin{center}
\leavevmode
\psfig{figure=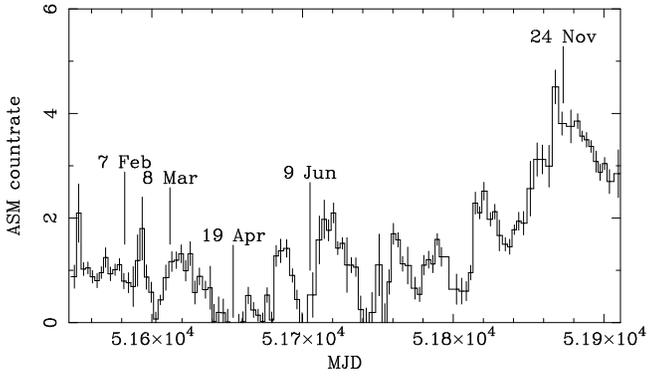,width=85truemm,angle=270}
\caption{The 3-day averaged {\em RXTE} 
All-Sky Monitor lightcurve for \lmc\ for
the year 2000, with the times of the \xmm\ observations marked.}
\label{fig:asm}
\end{center}
\end{figure}

Table \ref{tab:observations} lists the \xmm\ observations of \lmc\ that will be
used in this analysis, and in Fig. \ref{fig:asm} the observation times are 
marked on the {\em RXTE} All-Sky
Monitor lightcurve for \lmc\ during 2000. 
\lmc\ was also observed by \xmm\ in revolution 0028 on
the 2nd February 2000, but we have excluded this observation from the present
analysis because we are unable to produce a satisfactory effective area
calibration for rev. 0028, in which the source is off-axis. Over the ten months
spanned by these observations \lmc\ varied by more than a factor of 1000 in
RGS count rate.

\begin{table}
\caption{XMM observations of \lmc}
\label{tab:observations}
\begin{tabular}{lcccc}
Date&XMM&Instrument&Exposure$^{*}$&Count rate$^{*}$\\
(2000)&Orbit&&(ks)&(count s$^{-1}$)\\
\hline
07 Feb &0030&RGS1  &12.1&7.1\\
08 Mar &0045&RGS1+2&20.8&5.9\\
19 Apr &0066&RGS1+2&44.3&0.013\\
09 Jun &0092&RGS1+2&77.5&2.2\\
24 Nov &0176&RGS1+2&20.9&18.0\\
&&&&\\
\multicolumn{5}{l}{$^{*}$ Where applicable this is the mean of RGS1 and
RGS2}\\
\end{tabular}
\end{table}

\begin{figure*}
\begin{center}
\leavevmode
\psfig{figure=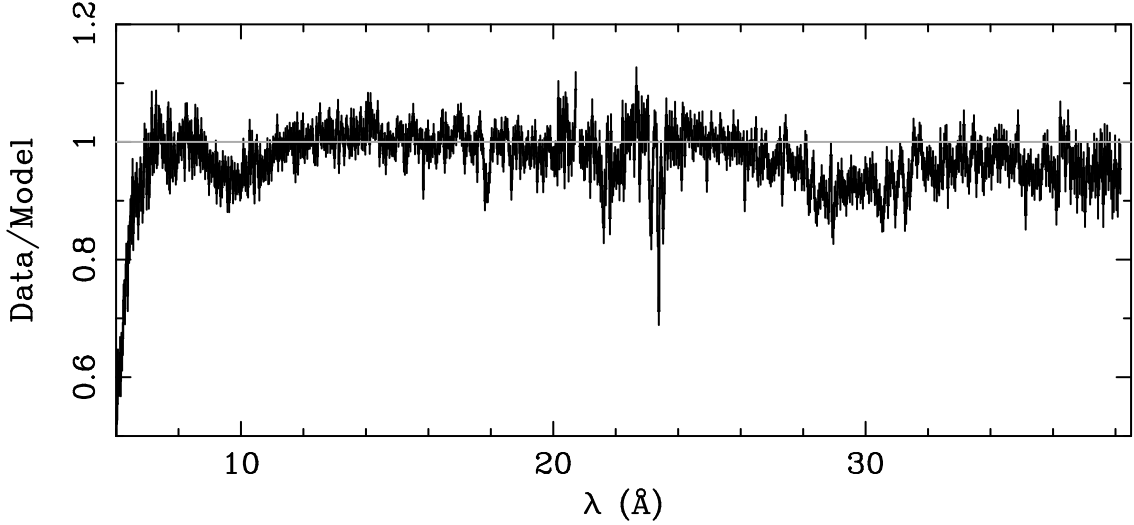,width=150truemm,angle=-90}
\caption{Data/model ratio for the RGS spectrum of Mrk~421 used to correct
residual artifacts in the effective area calibration.}
\label{fig:corr_factor}
\end{center}
\end{figure*}

The data were initially processed using the public release \xmm\ SAS
version 5.2. 
First and second order spectra from both RGS were
extracted (except for revolution 0030 in which only RGS1 was
operating) and response matrices were generated using the latest
version of the SAS task {\small RGSRMFGEN} 
(corresponding to SAS version 5.3.3).
To correct for residual artifacts in the effective
area calibration (see den Herder \nocite{denherder03} 2002), 
the effective area of each response
matrix was then divided by the ratio of a powerlaw + Galactic column
fit to the \xmm\ rev 0084 RGS spectra of the continuum source Mrk~421 
(Brinkmann \etal \nocite{brinkmann01} 2001).
\lmc\ was on-axis for all of the observations listed
in Table \ref{tab:observations}, as was Mrk 421 in revolution 0084.
 The data/model ratio of the
Mrk~421 RGS spectrum is shown in Fig. \ref{fig:corr_factor}; for most of the
RGS wavelength range the effective area correction is only a few percent.
The statistical errors on effective area correction were added in 
quadrature to the
statistical errors on the data.
Finally, for each revolution, spectra and response matrices were resampled
and coadded 
to produce a single spectrum per 
observation. To improve signal to noise, spectra from the February, March, June
and November observations were grouped by a factor 
of 3, resulting in spectra with $\sim 1000$ channels $\sim 30$ m\AA\ wide, 
well sampled with respect to the RGS resolution of $\sim 70$ m\AA\ FWHM.
The flux level of the April spectrum is extremely low, neccessitating very
heavy binning before a respectable signal to noise ratio is achieved; we have
grouped this spectrum by a factor of 64.
The resultant spectra were then analysed using the XSPEC software package
version 11.2.0 (Arnaud \nocite{arnaud96} 1996).

\section{Results}

\begin{figure}
\begin{center}
\leavevmode
\psfig{figure=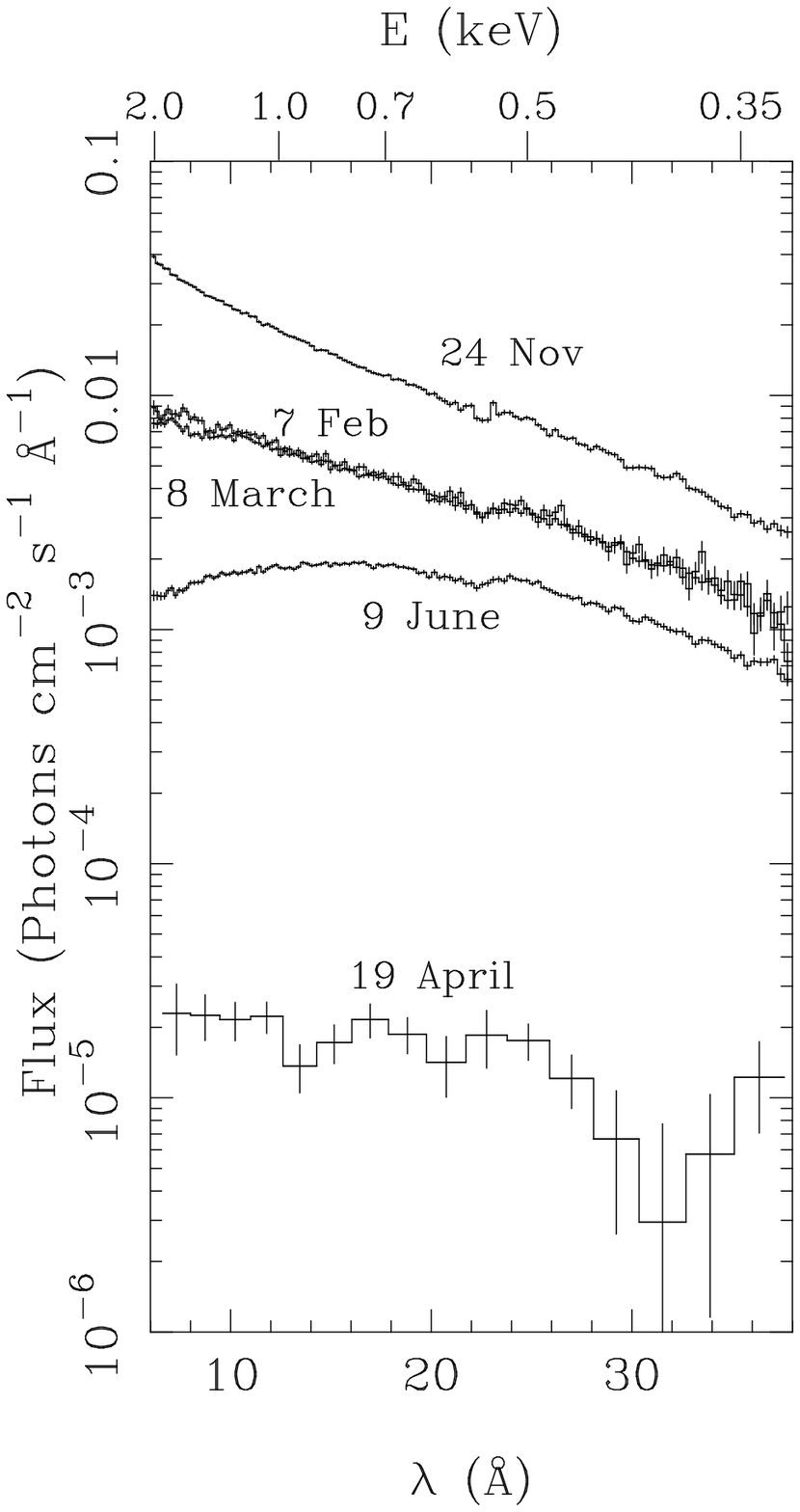,width=80truemm}
\caption{The RGS spectra of LMCX3 from the five different observations. To
improve presentation the four brightest spectra have been rebinned by a factor
of 8.}
\label{fig:allspec}
\end{center}
\end{figure}

Fig. \ref{fig:allspec} shows the flux calibrated spectra from all 
five observations. 
The four brightest spectra are shown at full resolution in Figs 
\ref{fig:0030spec} - \ref{fig:0176spec}. The spectra appear to be continuum 
dominated without prominent absorption or emission lines; this is particularly
clear in the 24th November spectrum, which has the highest signal to noise
ratio and shows no significant spectral features except for the interstellar 
O~I edge at 23\AA. 

\subsection{X-ray absorption in LMC X-3}

The degeneracy between absorption and the intrinsic spectral shape has been a
major source of uncertainty in attempts to model relatively low 
resolution soft
X-ray spectra of BHCs (e.g. Haardt \etal \nocite{haardt01} 2001). 
The excellent combination of sensitivity and resolution
afforded by the \xmm\ RGS allows individual absorption edges and lines to be
measured more or less independently of the underlying continuum shape, allowing
a far more reliable determination of the amount of soft X-ray absorption.

\subsubsection{Cold material}
\label{sec:cold}

\begin{figure*}
\begin{center}
\leavevmode
\psfig{figure=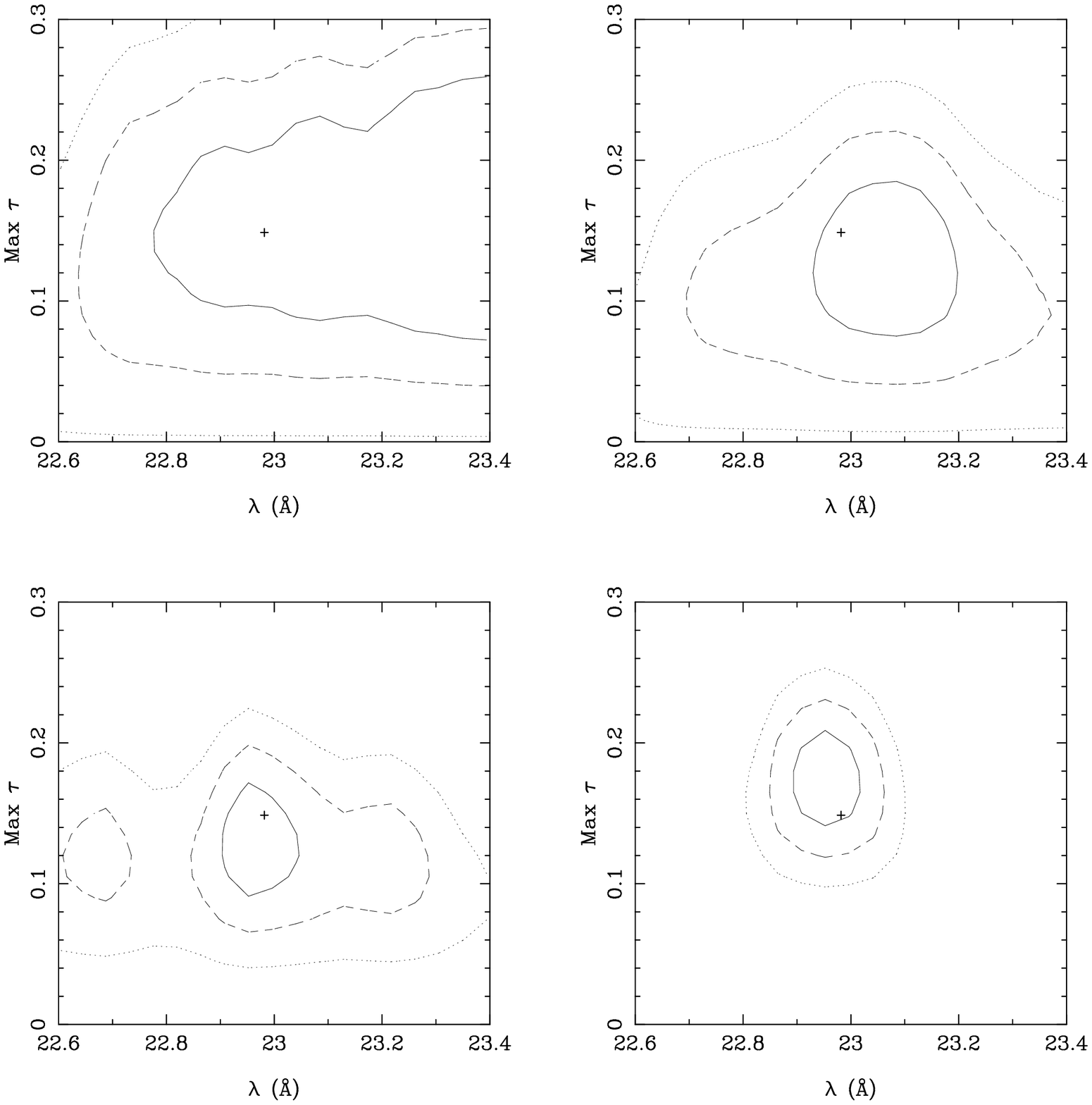,width=150truemm}
\caption{Confidence intervals for the depth and wavelength of the O~I K edge
for the February, March, June and November observations. The solid, dashed, and
dotted lines correspond to 1,2 and 3$\sigma$ respectively. The best fit edge
parameters from a simultaneous fit to all four spectra is marked with a cross
in each panel.}
\label{fig:pocontedge}
\end{center}
\end{figure*}

Cold material (either interstellar or local to the source) imprints a well
known sequence of absorption edges on X-ray spectra. For cosmic abundances,
oxygen produces by far the strongest absorption edge in the RGS energy range,
at 23~\AA, and inspection of Fig. \ref{fig:allspec} shows that this is
consistently the strongest feature observed in the RGS spectra of \lmc.  We
have therefore used the depth of the 23~\AA\ O~I K edge to determine the cold
absorbing column for each of the spectra. By studying a small spectral region
(3\AA\ to either side of the edge) we ensure that our column density
determination is essentially independent of the overall continuum shape.
The confidence intervals for the position of the edge and its optical depth are
shown in Fig. \ref{fig:pocontedge} for all the observations excluding the very
faint April observation. We have also fit the four observations simultaneously
with a single set of edge parameters but with the continuum slope and 
normalisation allowed to vary independently for each observation. The best fit
value for the simultaneous fit is shown as a cross on the confidence 
intervals in
Fig. \ref{fig:pocontedge}. The edge parameters derived from each of the
observations independently are consistent with the best fit edge parameters
from the simultaneous fit; there is no significant change in the column 
density of cold material from one observation to the next. The wavelength of
the edge from the simultaneous fit is $\lambda=22.98^{+0.09}_{-0.07}$\AA\ 
where the
confidence interval is 95\%. This is consistent with the wavelength
of 
the edge calculated by Verner \etal (1996) \nocite{verner96}, which is the 
edge wavelength
used in the {\small XSPEC} `tbabs' model (Wilms, Allen \& 
McCray \nocite{wilms00} 2000). 
To determine the equivalent H~I column density we 
substituted the `tbabs' 
model for the edge and performed another simultaneous fit
to the four spectra. This results in ${\rm N_{H}} = 3.8^{+0.8}_{-0.7} \times
10^{20}$ cm$^{-2}$, which is consistent with the H~I column density in the
direction of \lmc\ as
determined from radio observations ($3.2 \times 10^{20}$ cm$^{-2}$; Nowak et
al. \nocite{nowak01} 2001).

\begin{figure*}
\begin{center}
\leavevmode
\psfig{figure=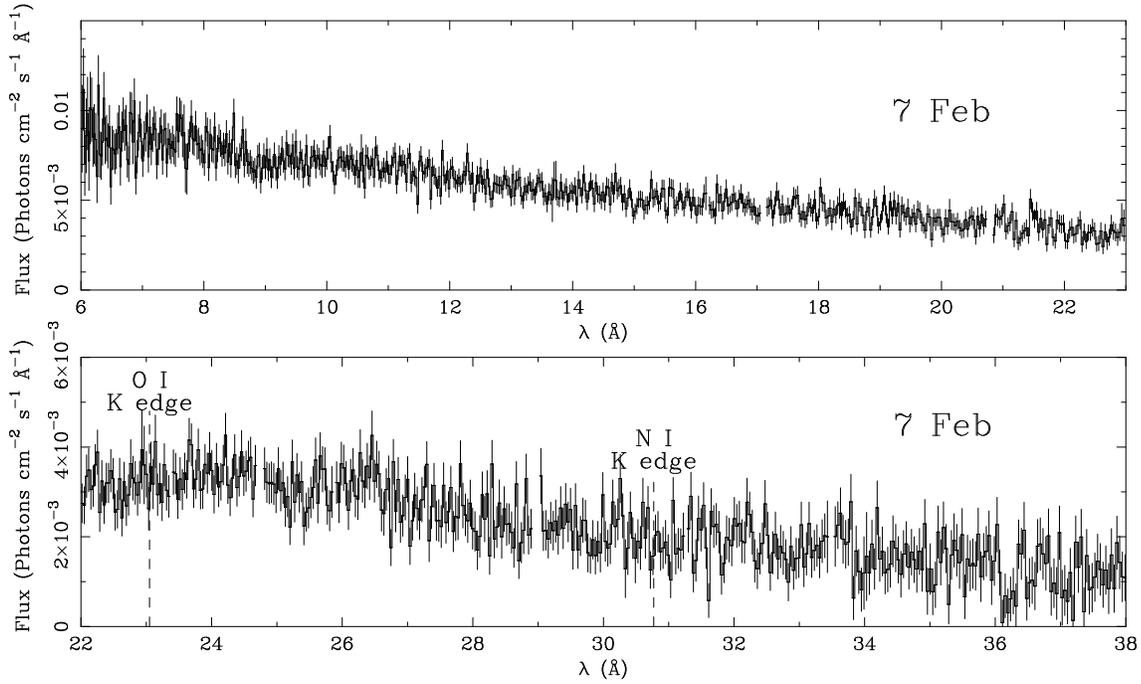,width=150truemm,angle=270}
\caption{RGS spectrum from 07 February 2000 }
\label{fig:0030spec}
\end{center}
\end{figure*}

\begin{figure*}
\begin{center}
\leavevmode
\psfig{figure=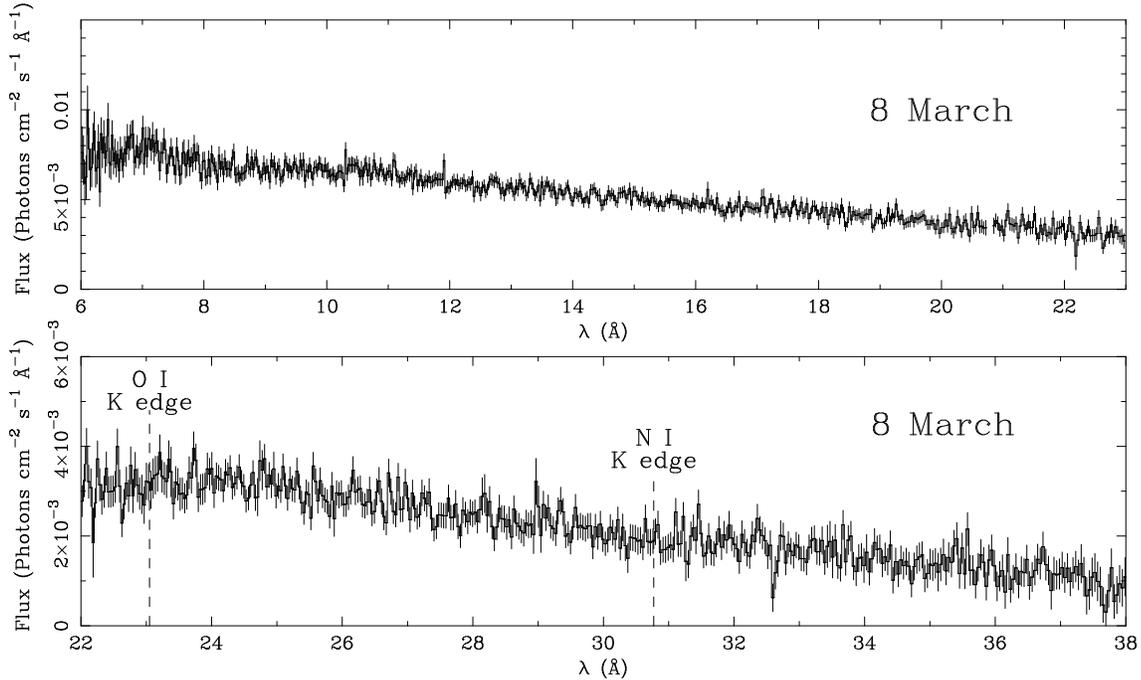,width=150truemm,angle=270}
\caption{RGS spectrum from 08 March 2000 }
\label{fig:0045spec}
\end{center}
\end{figure*}

\begin{figure*}
\begin{center}
\leavevmode
\psfig{figure=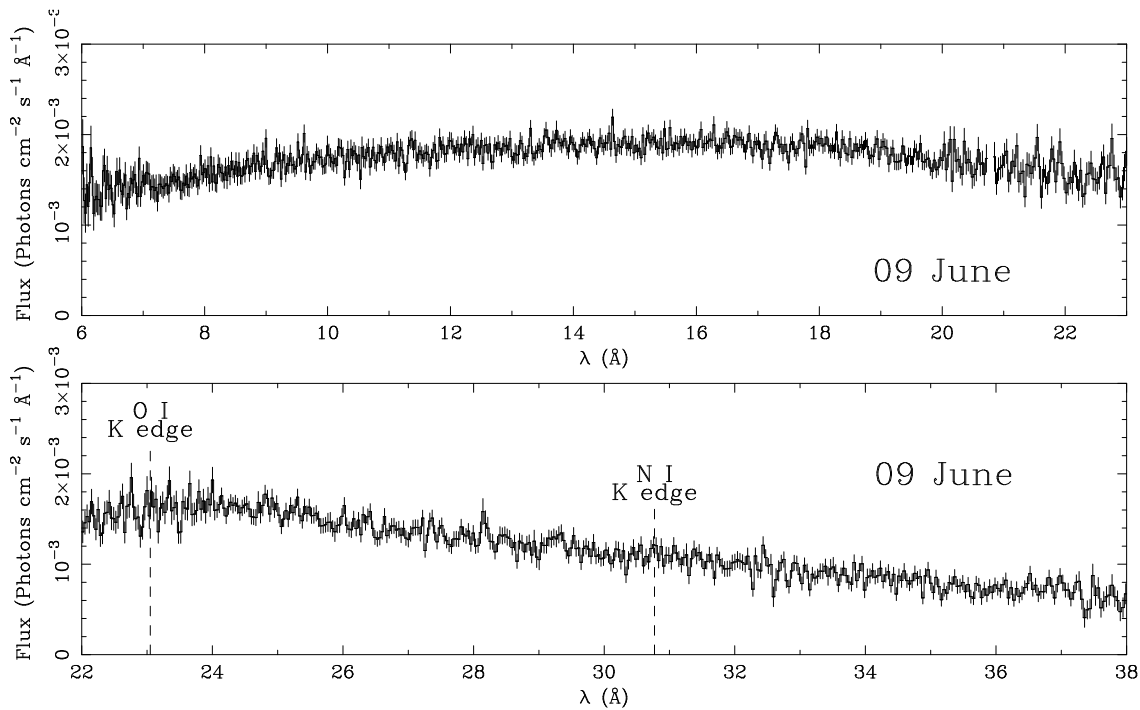,width=150truemm,angle=270}
\caption{RGS spectrum from 09 June 2000}
\label{fig:0092spec}
\end{center}
\end{figure*}

\begin{figure*}
\begin{center}
\leavevmode
\psfig{figure=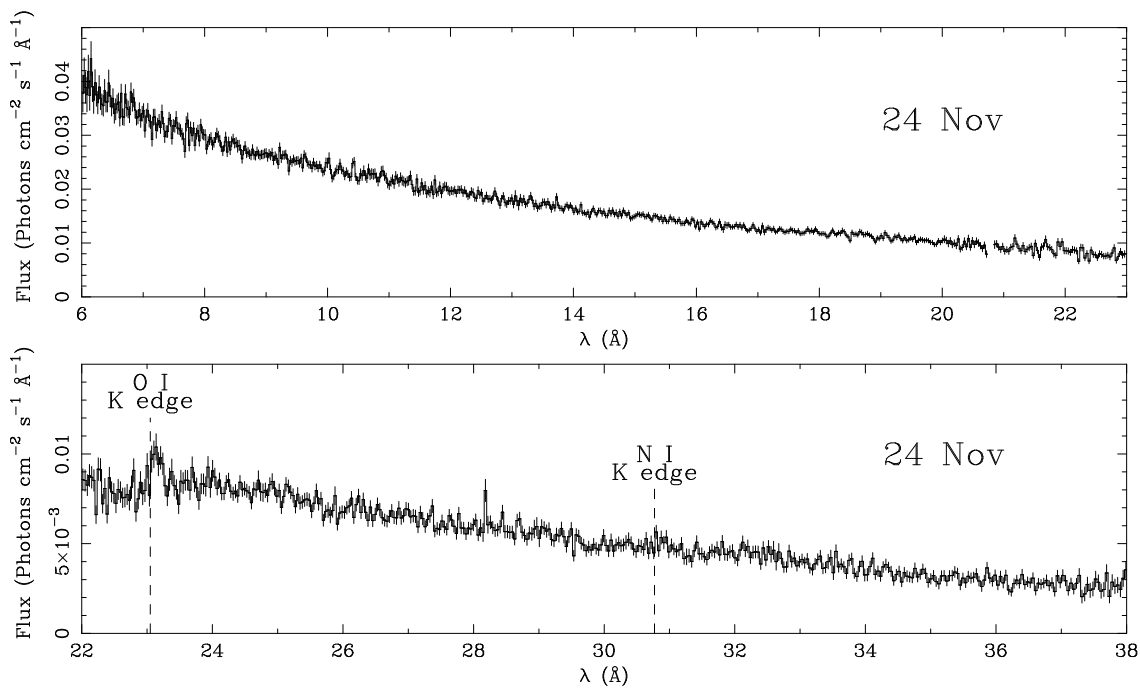,width=150truemm,angle=270}
\caption{RGS spectrum from 24 November 2000}
\label{fig:0176spec}
\end{center}
\end{figure*}

\subsubsection{Ionized material}
\label{sec:ionized}

Although the column density of cold absorbing material is consistent with the 
interstellar medium of the Galaxy, absorbing material associated with a wind
from the accretion disk or the secondary star may be highly ionized by the 
intense X-ray and ultraviolet radiation. Absorption and emission lines from 
warm, ionized
material have now been detected in the soft X-ray band from a number of AGN 
and X-ray binaries (e.g. Kaastra \etal \nocite{kaastra00} 
2000, Brandt \& Schultz \nocite{brandt00} 2000).

There are no obvious emission or absorption lines present in the RGS spectra of
\lmc\ (Figs \ref{fig:0030spec}, \ref{fig:0045spec}, \ref{fig:0092spec} and \ref
{fig:0176spec}) and so we have attempted a sensitive search for their presence.
The interstellar medium and the Local-Group intergalactic medium have imprinted
absorption lines of O~I at 23.52\AA\ and O~VII at 21.60\AA\ on the spectrum of
Mrk~421 (Paerels \etal \nocite{paerels03} 2003), which we have used to 
correct residual artifacts in the
effective area calibration. 
Therefore we
performed the absorption line search on spectra which have no effective
area correction. 
First, we calculated the transmission through the interstellar medium 
as a function of energy using `tbabs' with ${\rm N_{H}} = 3.8 \times
10^{20}$ cm$^{-2}$. We then corrected the spectra for interstellar 
absorption by dividing
the effective area of the response matrix by the transmission. We then produced
continuum spectra by smoothing our RGS spectra with a sliding box 
30 pixels ($\sim 1$\AA) wide. We rejected any points more than 2$\sigma$
deviant from the resulting smooth curve (i.e. potential emission/absorption
features), and repeated the smoothing. This smoothing/rejection process was
iterated several times to ensure that even marginally significant discrete
features do not influence its shape.  We then performed a grid search for
emission and absorption lines where the grid was defined in terms of central
wavelength, FWHM and equivalent width. The grid values of central wavelength
were chosen to oversample the wavelength bins of the data, FWHM was stepped
between 0 and 1000 km s$^{-1}$, with a stepsize of
200 km s$^{-1}$ (well sampled at the RGS resolution), and the stepsize of the 
equivalent
width was matched to the statistical uncertainty on the data.
At each point in the grid, we formed a 
model
spectrum by adding a gaussian 
emission/absorption line to the smooth continuum spectrum, and we computed the
goodness of fit of this model to the real (unsmoothed) spectrum using the
poisson errors on the data points and the $\chi^{2}$ statistic. The
significance of any features, and limits on the equivalent width were
determined using $\Delta \chi^{2}$. Since we search for lines in either 
emission 
or absorption, we expect $\Delta \chi^{2}$ to give an approximate measure of
the significance of any lines (see Protassov et al. \nocite{protassov02} 
2002).  
The grid search was performed 
using our own software which propogates the model through the response matrix
in an identical fashion to {\small XSPEC}. 

\begin{figure}
\begin{center}
\leavevmode
\psfig{figure=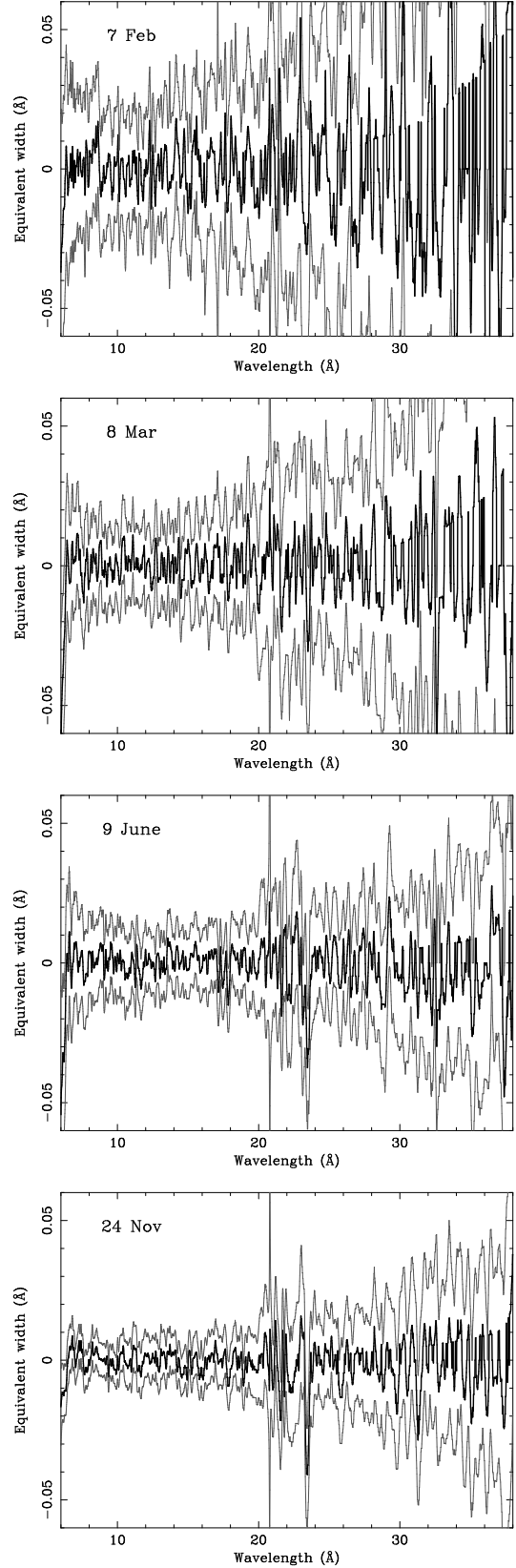,width=70truemm}
\caption{Equivalent widths of emission and absorption lines of up to 1000 km 
s$^{-1}$
FWHM from the line
search described in Section \ref{sec:ionized}; negative equivalenth width
refers to absorption lines. The black curve shows the best
fit equivalent width as a function of wavelength; the grey curves show the
$3 \sigma$ ($\Delta \chi^{2} = 9$) upper and lower limits.}
\label{fig:ewlimits}
\end{center}
\end{figure}

The results of this line search are shown in Fig. \ref{fig:ewlimits}. 
There are $\sim$1000 channels in each spectrum, and therefore we expect
$\sim 1$ spurious feature at the $ 3\sigma$ level 
in each spectrum. The spectra from the 7th February and 8th March each 
contain one 
$3\sigma$ significant absorption feature, at 36.18$\pm0.08$\AA\ and 
32.61$\pm0.05$\AA\
respectively. These do not correspond to the wavelengths of strong transitions
in abundant elements, and are therefore probably spurious. In the spectrum 
from 9th June, there are three $3\sigma$ significant
absorption features, at 17.86$\pm0.08$\AA, 21.62$\pm0.04$\AA\ and 37.38$\pm0.05$\AA, and one $5\sigma$
significant feature at 23.51$\pm0.03$\AA. The features at 17.86\AA\ and 
37.38\AA 
do not correspond to the wavelengths of any expected absorption line and are
again probably spurious. The 21.62\AA\ feature has an equivalent width of
19$\pm 12$m\AA\ and 
is consistent with the $1s - 2p$
transition in O~VII, while the 23.51\AA\ feature is the $1s - 2p$
transition of O~I. 
In
the spectrum from 24th November, the most significant feature ($6 \sigma$) 
is again the $1s - 2p$ transition in O~I. 
In
addition, there
are three $3\sigma$ and one $4\sigma$ significant absorption features in this 
spectrum. Two of
the $3\sigma$ features, at
14.94$\pm 0.04$\AA\ and 31.30$\pm 0.06$\AA\ do not correspond to strong
transitions in abundant elements and are therefore probably spurious. The 
other $3\sigma$ line, at 
$18.93\pm
0.05$\AA, corresponds to O~VIII Ly$\alpha$, while the $4\sigma$ line is
found at $21.58\pm 0.03$\AA, and corresponds to O~VII He-$\alpha$. 
The 18.93\AA\
and 21.58\AA\ have equivalent widths of $9\pm4$m\AA\ and $21\pm12$m\AA\
respectively. 

The O~VII lines detected on the 9th June and 24th November have similar
equivalent widths of $\sim$20m\AA, which implies an O~VII column density 
between 3$\times
10^{15}$~cm$^{-2}$ and 5$\times 10^{16}$~cm$^{-2}$. Intervening material
associated with the halo of our Galaxy is likely to make a significant
contribution to (and may dominate) this line, because Galactic O~VII
absorption is seen with a similar equivalent width in the blazars observed 
during
RGS calibration (Paerels \etal \nocite{paerels03} 2003). It is notable that 
if we apply the Mrk~421 effective area
correction before performing the line search on \lmc, the OVII absorption lines
are not detected in any of the spectra. 
The equivalent width of the O~VIII Ly$\alpha$ absorption observed on the 24th 
November implies an O~VIII column density of between 4$\times
10^{15}$~cm$^{-2}$ and 1.4$\times
10^{16}$~cm$^{-2}$. This absorption line is seen in two out of three RGS
calibration blazars (Paerels \etal \nocite{paerels03} 2003) with a similar
equivalent width, and so the line in \lmc\ may or may not arise in the Galactic
halo. The observed wavelength is consistent with an outflow from \lmc, but is
not well enough constrained to rule out a Galactic origin.

We can place very firm upper limits on the column density of an ionized wind by
using the 3$\sigma$ upper limits on the equivalent widths of $1s - 2p$
absorption lines from O~II -- O~VIII obtained from Fig. \ref{fig:ewlimits} and
assuming that there is no interstellar component to these lines.
The spectra from 9th June and 24th November both have sufficient signal to
noise ratios to provide useful limits. These limits are given in Table
\ref{tab:ewlimits}, along with the limits on column density assuming
a velocity width $\sigma > 100$ ${\rm km s^{-1}}$. The inferred total column 
density for O~II -- O~VIII is $<5.2 \times 10^{17} {\rm cm^{-2}}$ on
the 9th June and $<2.5\times 10^{17} {\rm cm^{-2}}$ on the 24th November.

\begin{table}
\caption{3$\sigma$ upper limits to equivalent widths of O~II -- O~VIII 
$1s - 2p$
absorption lines. The upper limits to column density have been calculated
assuming velocity width $\sigma > 100$~${\rm km s^{-1}}$.}
\label{tab:ewlimits}
\begin{tabular}{lccccc}
&&\multicolumn{2}{c}{--- 9 Jun ---}&\multicolumn{2}{c}{--- 24 Nov ---}\\
Ion & $\lambda$ & EW & $N$ & EW   & $N$ \\
   &(\AA)& (m\AA) & ($10^{16}$cm$^{-2}$) & (m\AA) &($10^{16}$cm$^{-2}$) \\
\hline
O~VIII&18.97&$<8$ &$<0.8$ &$<17$ &$<2.2$\\
O~VII &21.60&$<37$  &$<11$ &$<36$ &$<9.0$\\
O~VI  &22.01&$<34$  &$<7.8$ &$<20$ &$<1.6$\\
O~V   &22.33&$<36$  &$<9.4$ &$<30$ &$<4.1$\\
O~IV  &22.78&$<13$  &$<2.4$ &$<23$ &$<5.3$\\
O~III &23.11&$<47$  &$<21$ &$<14$ &$<2.9$\\
O~II  &23.30&$<34$  &$<6.3$ &$<40$ &$<8.7$\\
\end{tabular}
\end{table}

\subsection{Continuum shape}
\label{sec:model}

Having established that the soft X-ray spectrum of \lmc\ is essentially free
of discrete emission and absorption features except for the interstellar
absorption, we now consider the shape of the continuum. For all the fits
performed in this section we include a cold absorbing column fixed at our 
measured value of $3.8 \times 10^{20}$ cm$^{-2}$, and all the results are
tabulated in Table \ref{tab:fits}.

\begin{figure*}
\begin{center}
\leavevmode
\psfig{figure=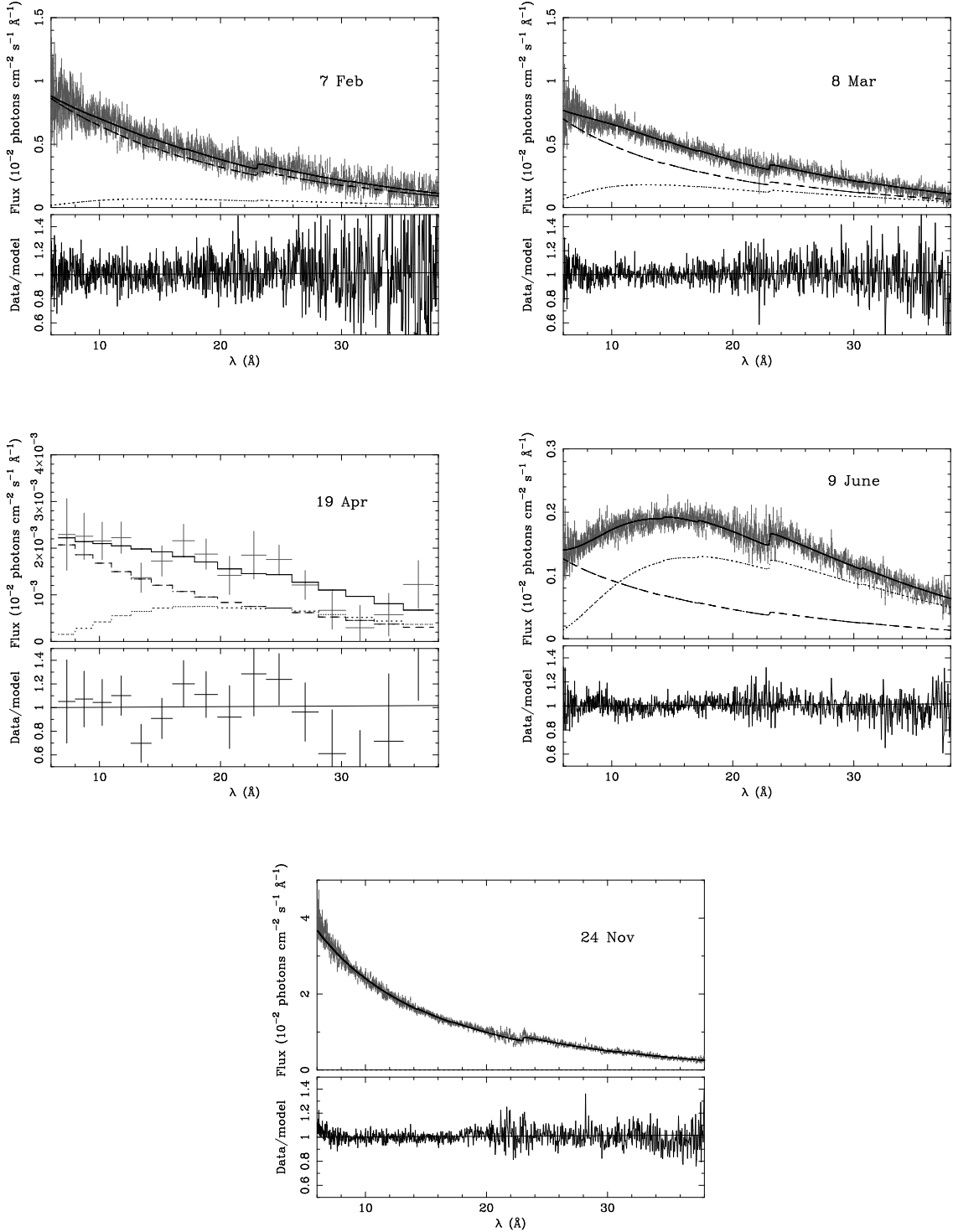,width=150truemm}
\caption{Best fit disk-blackbody + Comptonised disk models to the RGS
spectra. The dashed black line shows the disk-blackbody 
component, and the
dotted black line shows the Comptonised component; the solid line is the sum of
the two. The data are shown in grey. The data/model ratio is shown in the lower
panels; for clarity the error bars are not shown on the data/model ratio
plots except for the 19 April spectrum.}
\label{fig:diskbbcomp}
\end{center}
\end{figure*}

We started by fitting a disk-blackbody model to our spectra. Only the very
faint spectrum from 19th April is acceptably fitted 
by this model, which falls off
too rapidly below 10\AA\ to fit the other spectra. At shorter wavelengths, BHC
spectra typically require a powerlaw component in addition to (or instead of)
a disk blackbody (e.g. Ebisawa \etal \nocite{ebisawa96} 1996).  We therefore
added a powerlaw to the model, with the photon index constrained to lie within
the range of photon indices seen at higher energy in \lmc: $1.5 < \Gamma < 4.5$
(Wilms \etal \nocite{wilms01} 2001, Wu \etal \nocite{wu01} 2001). This resulted
in statistically acceptable fits to the February, March and April observations,
but not to the higher signal to noise June and November spectra. In all cases
the best fit photon index converged to the minimum of the acceptable
range, indicating that the powerlaw produces too many photons at long
wavelength, where the data are already quite well reproduced by the 
disk-blackbody component. Therefore we substituted the powerlaw component for
Compton scattered emission from a hot corona, making use of the {\small
THCOMPDS\footnote{THCOMPDS was provided by Piotr \.Zycki, and is available from
http://www.camk.edu.pl/$\sim$ptz/relrepr.html}}
model (\.Zycki, Done \& Smith \nocite{zycki01}
2001). This model upscatters photons to produce a power-law shape; the seed 
photons for the Compton scattered component were assumed to
come from the disk blackbody, and the electron temperature of the scatterer was
fixed to 100 keV (the RGS energy range is too limited for the fit to be 
sensitive to this parameter). This model produced better fits to all the
spectra, and all of the fits are acceptable at the 3$\sigma$ confidence level;
Fig. \ref{fig:diskbbcomp} shows RGS spectra compared to the best fit model
components. 

\begin{table*}
\caption{Model fit results. Errors are quoted at 95\% confidence for 1
interesting parameter. In cases for which a
parameter has reached a limit on the allowed range of values while 
$\Delta
\chi^{2} < 4$ the limit has been marked with an `*'. 
Fluxes are given in the 0.3
-- 2 keV RGS energy band}
\label{tab:fits}
\begin{tabular}{lcccccc}
\hline
\multicolumn{7}{c}{Disk blackbody}\\
&&&&&&\\
Observation & $kT_{\rm in}$ & Flux (disk) & - & - & $\chi^{2}/\nu$ & Prob\\
& (keV) & ($10^{-10}$ erg~s$^{-1}$~cm$^{-2}$) & & & & \\
&&&&&&\\
07 Feb & $0.63^{+0.02}_{-0.02}$ & $2.09^{+0.16}_{-0.13}$ & - & - & 1169/986 & $4.5 \times 10^{-5}$\\
08 Mar & $0.58^{+0.01}_{-0.01}$ & $1.99^{+0.09}_{-0.10}$ & - & - & 1161/997 & $2.4 \times 10^{-4}$\\
19 Apr & $0.41_{-0.07}^{+0.15}$ & $0.007^{+0.010}_{-0.005}$ & - & - & 15/16 & 0.53\\ 
09 Jun & $0.34^{+0.01}_{-0.01}$ & $0.68^{+0.02}_{-0.02}$ & - & - & 2872/998 & 0\\
24 Nov & $1.04^{+0.01}_{-0.01}$ & $6.72^{+0.24}_{-0.35}$ & - & - & 1344/1001 & $1.9 \times 10^{-12}$\\
&&&&&&\\
\hline
\multicolumn{7}{c}{Disk blackbody + powerlaw}\\
&&&&&&\\
Observation & $kT_{\rm in}$ & Flux (disk) & $\Gamma$ & Flux (powerlaw) & $\chi^{2}/\nu$ & Prob\\
& (keV) & ($10^{-10}$ erg~s$^{-1}$~cm$^{-2}$) & & ($10^{-10}$ erg~s$^{-1}$~cm$^{-2}$) & & \\
&&&&&&\\
07 Feb & $0.67^{+0.05}_{-0.02} $ & $1.25^{+0.30}_{-0.27}$ & $1.50_{-0.00*}^{+0.12}$ & $0.89^{+0.14}_{-0.20}$ & 1045/984 & 0.085 \\
08 Mar & $0.59^{+0.01}_{-0.01}$ & $1.30^{+0.12}_{-0.11}$ & $1.50_{-0.00*}^{+0.05}$ & $0.73^{+0.11}_{-0.08}$ & 954/995 & 0.82\\
19 Apr & $0.28^{+1.72*}_{-0.18*}$ & $0.005^{+0.002}_{-0.005*}$ & $1.50^{+3.50*}_{-0.00*}$ & $0.002^{+0.004}_{-0.002*}$ & 13/14 & 0.54\\ 
09 Jun & $0.28^{+0.01}_{-0.01} $ & $0.28^{+0.01}_{-0.01}$ & $1.50_{-0.00*}^{+0.02}$ & $0.43_{-0.03}^{+0.02}$ & 1132/996 & $1.7 \times 10^{-3}$\\
24 Nov & $1.18^{+0.04}_{-0.04} $ & $6.04^{+0.31}_{-0.29}$ & $1.50_{-0.00*}^{+0.09}$ & $0.73_{-0.14}^{+0.08}$ & 1210/999 & $4.7 \times 10^{-6}$\\
&&&&&&\\
\hline
\multicolumn{7}{c}{Disk blackbody + Comptonised disk (thcompds)}\\
&&&&&&\\
Observation & $kT_{\rm in}$ & Flux (disk) & $\Gamma$ & Flux (Comptonised) & $\chi^{2}/\nu$ & Prob\\
& (keV) & ($10^{-10}$ erg~s$^{-1}$~cm$^{-2}$) & & ($10^{-10}$ erg~s$^{-1}$~cm$^{-2}$) & & \\
&&&&&&\\
07 Feb & $0.32_{-0.07}^{+0.05}$ & $0.23_{-*}^{+0.14}$ & $1.60_{-0.10*}^{+0.38}$ & $1.92^{+0.29}_{-0.29}$ & 1024/984 & 0.18 \\
08 Mar & $0.36^{+0.03}_{-0.03}$ & $0.61_{-0.42}^{+0.07}$ & $1.50_{-0.00*}^{+0.42}$ & $1.42_{-0.13}^{+0.49}$ & 902/995 & 0.98 \\
19 Apr & $0.24_{-0.14*}^{+0.24}$ & $0.003_{-0.003*}^{+0.006}$ & $1.50^{+3.50*}_{-0.00*}$ & $0.005_{-0.005*}^{+0.003}$ & 12/14 & 0.59\\ 
09 Jun & $0.26^{+0.01}_{-0.01}$ & $0.43\pm 0.02$ & $1.50_{-0.00*}^{+0.33}$ & $0.28_{-0.01}^{+0.04}$ & 1079/996 & 0.034\\
24 Nov & $0.55_{-0.01}^{+0.02}$ & $0.0_{-0.0*}^{+0.1}$ & $1.50_{-0.00*}^{+0.03}$ & $6.76_{-0.16}^{+0.16}$ & 1127/999 & $2.8 \times 10^{-3}$\\
&&&&&&\\
\hline
\end{tabular}
\end{table*}

\section{Discussion}

In the preliminary analysis of \xmm\ EPIC and RGS spectra, Wu \etal
\nocite{wu01} (2001) argued that mass transfer mainly 
occurs in \lmc\ by Roche lobe
overflow rather than via a strong stellar wind because the line of sight
absorption is small ($\nh < 10^{21}$ cm$^{-2}$). 
With the more detailed analysis
presented here we can rule out accretion from a stellar wind with much higher
confidence.

For a continuous spherical wind leaving a body of mass $M$ and radius $R$ at
the escape velocity, the hydrogen column density $\nh$ is related to the mass-loss rate by:
\begin{equation}
\frac{dM}{dt} = 24.7 m_{p} \sqrt{GMR} \times \nh
\label{eq:massloss}
\end{equation}
where $m_{p}$ is the mass of a proton. 
This expression can be used to obtain a conservative upper limit to
the wind mass-loss rate, because if we assumed a 
$\beta$-law velocity
profile ($dr/dt \propto [1-R_{*}/r]^{\beta}, r > 1.04 R_{*}$), 
as used to model the 
line-driven winds from hot stars
(Pauldrach, Puls \& Kudritzki \nocite{pauldrach86} 1986), instead of a
constant velocity wind, the mass-loss rate
for a given column density would be lower than that given in equation
(\ref{eq:massloss}).
In Section \ref{sec:cold} we obtained an upper limit of $1.4 \times 10^{20}$
cm$^{-2}$ for the hydrogen column density intrinsic
to \lmc, based on the neutral O edge and assuming
Solar metallicity.  
If we instead assume a metal abundance of 0.4 $\times$ Solar as
expected in the LMC (Caputo \etal \nocite{caputo99} 1999), we find an
upper limit of 
 $\nh < 3.5\times 10^{20}$ cm$^{-2}$. We can also estimate the hydrogen column 
density
associated with O~II -- O~VIII using the limits obtained in Section
\ref{sec:ionized} from the 24th November spectrum. Assuming 0.4 $\times$ Solar
abundance, we obtain $\nh < 7.4\times 10^{20}$
cm$^{-2}$. Summing these two upper limits to obtain the total hydrogen column
density associated with O~I -- O~VIII, we find $\nh < 1.1\times
10^{21}$ cm$^{-2}$.
Assuming that the companion star has mass $M < 7 M_{\sun}$ and radius $R < 5
R_{\sun}$ (Soria \etal \nocite{soria01} 2001) and substituting into equation 
(\ref{eq:massloss}), we find that $dM/dt < 8 \times 10^{17}$ g
s$^{-1}$. This
is insufficient to power the X-ray emission except when \lmc\ is
in its very weakest state (e.g. April 2000). 

However, we have so far neglected
wind material in which O is completely ionized. When the luminosity of the 
system is as high as $10^{38}$ erg s$^{-1}$, the ionization parameter
($\xi = L/nr^{2}$) at the 
base of a homogeneous,
constant-velocity wind from the secondary star 
is $\sim 3 \times 10^{25}
/ \nh$ where $L$ is luminosity in erg s$^{-1}$, $n$ is the ion density in
cm$^{-3}$, and 
$r$ is the binary separation, $\sim 10^{12}$ cm. Thus for mass-loss
rates $< 3 \times 10^{20}$~g~s$^{-1}$ 
we expect $\xi > 100$~erg~cm~s$^{-1}$, 
implying that the majority of O in the wind is completely ionized
(Kallman \& McCray \nocite{kallman82} 1982), and
therefore undetectable by X-ray absorption line spectroscopy. For a wind with a
$\beta$-law velocity
profile the wind is much denser near the base, but at a distance of 
$R_{*}/4$
from the stellar photosphere, when the wind has only achieved about a fifth of 
its terminal
velocity (assuming $\beta \sim 1$), $\xi$ reaches $\sim 10^{25} / \nh$.
A stellar wind cannot be driven by line-pressure at such a
high ionization state, because 
the ions that provide most of the acceleration (Abbot \nocite{abbot82} 1982) 
will not be present (McCray \& Hatchett \nocite{mccray75} 1975). On the other
hand, we can rule out a mass-loss rate of $> 3 \times 10^{20}$~g~s$^{-1}$,
because the column
density would be high enough, (and $\xi$ low enough) that absorption lines
from O would be very apparent in the RGS spectra. Thus the lack of absorption 
lines
in the RGS spectra implies that a line-driven stellar wind is not
the primary source of accretion material for \lmc\ except, possibly, when it is
in a faint state. By comparing the variability properties of \lmc\ to LMC X-1, 
Nowak \etal \nocite{nowak01} (2001) also conclude that \lmc\ is not accreting
from a stellar wind. Roche lobe overflow is therefore the most likely mode 
of mass transfer in
\lmc.

In order to explain the state changes observed in \lmc\ along with the long
term $\sim 100$ day periodic variability, Wilms \etal \nocite{wilms01} (2001)
propose that the accretion rate could be controlled by an accretion disk
wind-driven limit cycle as discussed by Shields \etal \nocite{shields86}
(1986). The optical photometric variability presented by Brocksopp \etal
\nocite{brocksopp01} (2001) is also consistent with this hypothesis. In this 
scenario, the central X-ray source in \lmc\ drives a Compton
heated wind from the accretion disk that is sufficiently strong that it
interrupts its flow of fuel. For such an instability to take place, the mass
loss in the wind
must exceed the accretion rate by a considerable factor 
(e.g. $\geq 30$, Shields \etal \nocite{shields86} 1986). 
At distances from the central source greater than the
outer radius of the accretion disk, the Compton heated wind is approximately
spherical (Begelman \& McKee \nocite{begelman83b} 1983) and so we can use
equation \ref{eq:massloss} to relate the mass loss rate of an accretion disk
wind to the column density.  Taking the radius of the accretion disk to be
$\sim 10^{12}$ cm, and the mass of the primary to be $\sim 10 M_{\sun}$ (Soria
\etal \nocite{soria01} 2001), we find that an accretion disk wind capable of
driving the long-term variability will have $\nh
> 3 \times 10^{22}$ cm$^{-2}$ when the luminosity of the source is $\sim
10^{38}$ erg s$^{-1}$. If we assume a wind temperature of $\sim 10^{8}$~K
(Begelman, McKee \& Shields \nocite{begelman83a} 1983), more than $99.9\%$ of O
is in the form of O~IX, and the column density of O~VIII can therefore be $<
10^{16}$ cm$^{-2}$, consistent with the upper limits given in Table
\ref{tab:ewlimits}. Thus we cannot rule out that the periodic variability in
\lmc\ is driven by an accretion disk wind.
However the column density in any accretion disk wind must be small enough 
that that it is transparent
to scattering: thermal emission from a 
wind with column density $>10^{24}$ cm$^{-2}$ would dominate the X-ray 
luminosity of 
the system, giving rise to prominent
K$\alpha$ lines from Fe~XXV and Fe~XXVI. 
This is not observed in \lmc\ (Wu \etal
\nocite{wu01} 2001). Thus even if there is an accretion disk wind, it imparts 
no significant features or
distortions on the X-ray spectrum of \lmc. 

The resolution and throughput of the RGS coupled with the lack of X-ray 
absorption in \lmc\ provide a clearer picture of the soft X-ray continuum 
spectral shape than has ever been possible before. Excluding the very faint
April spectrum, the spectral fitting
performed in Section \ref{sec:model} demonstrates that a single disk blackbody
is unable to describe the 0.3-2 keV spectral shape even when \lmc\ is in the 
high/soft state, mainly because the model falls off too fast at short
wavelength. Adding a powerlaw component results in much improved
(though for the 24th November still unacceptable) fits. However, the fitted 
powerlaw
slopes consistently bottom out at the hardest allowed values in the fitting 
($\Gamma=1.5$) and are considerably harder than the powerlaw slopes found over
the 0.3-12 keV energy range with \xmm\ EPIC (Wu \etal \nocite{wu01} 2001), 
even when the EPIC and RGS data were obtained simultaneously. 

A visual
inspection of Fig. \ref{fig:allspec} offers some insight into the reason for
the discrepant powerlaw parameters: longward of 23 \AA\ the spectra have
almost the same shape, running approximately straight and parallel as expected 
from a disk blackbody with a varying inner temperature. 
However, a powerlaw, even with a photon index as hard as $\Gamma=2$, has a
softer spectral shape than a disk blackbody longward of the peak. Thus for a
powerlaw with $\Gamma > 2$ to make a significant contribution at $\lambda <
10$\AA\ it would have to make a larger contribution at longer wavelength, 
and
therefore the spectra should deviate significantly from the disk blackbody
shape at  $\lambda > 23$\AA.  
Therefore the straight parallel spectra at $\lambda > 23$\AA, and the low
best-fit photon indices, imply that the component which is observed 
to have a powerlaw shape at $< 6$ \AA\ (i.e. at E $> 2$ keV) does not continue
with the same powerlaw shape through the soft X-ray band. 

The powerlaw
component is normally assumed to be the Compton upscattered emission from a
hot corona (Sunyaev \& Titarchuk \nocite{sunyaev80} 1980). 
The Compton scattered spectrum is
expected to  deviate from a powerlaw shape close to the energy of the seed 
photons. Thus the absence of the
powerlaw component in the soft X-ray is consistent with the seed photons for
the Comptonisation coming from the disk blackbody component. This is 
predicted by
successful models for black hole binary spectral states (e.g. Haardt \&
Maraschi \nocite{haardt91} 1991, Haardt \&
Maraschi \nocite{haardt93} 1993, Chakrabarti \&
Titarchuk \nocite{chakrabarti95} 1995, 
Haardt \etal \nocite{haardt97} 1997), in which the temperature and optical
depth of the corona depend on the amount of cooling via Compton scattering off
soft X-ray photons from the disk.
This result is borne out by the improved fits which are obtained when the 
powerlaw is
exchanged for a Comptonised disk blackbody. However, even with this model, the
photon index which describes the Comptonised spectrum towards shorter
wavelengths is lower than the photon indices obtained over a larger energy
range using EPIC. Furthermore, for the brightest spectra the soft X-ray flux
from the Comptonised component dominates the flux directly emerging from the 
disk. This suggests some further complexity in the spectral shape at the
short wavelength end of the spectra (the disk blackbody and Comptonised 
components have a similar spectral
shape at the long wavelength end). One possible explanation could be that
in the soft state, the soft component in \lmc\ falls off slower than a disk
blackbody towards short wavelengths due to electon scattering near the surface
of the disk.
Such a broadening of the soft component spectral shape has been observed in the
soft X-ray transients (e.g. GS 2000+25, GRO 1655$-$40) studied by \.Zycki \etal
\nocite{zycki01} (2001) using {\em Ginga} and {\em RXTE} spectra. 
However, the RGS does not extend to short enough wavelengths for a
proper investigation of this, which we defer to a future study using the EPIC
instruments which cover a wider energy range.  

\section{Conclusions}

We have presented \xmm\ RGS spectra of \lmc\ covering the full range of
spectral states observed in this source (more than 3 orders of magnitude in RGS
count rate). The spectra are completely dominated by continuum emission, and are
devoid of notable spectral features except for the interstellar O~I edge. The
depth of the edge is consistent between the different observations, and is
consistent with the Galactic interstellar column density of $3.2 \times 10^{20}
{\rm cm^{-2}}$ in the direction of \lmc. The upper limit to the edge depth
implies $\nh < 4 \times 10^{20} {\rm cm^{-2}}$ associated with neutral O,
intrinsic to \lmc.  Upper limits on the equivalent widths of $1s - 2p$ resonant
absorption lines from O~II -- O~VIII imply $\nh < 7.4 \times 10^{20} {\rm
cm^{-2}}$ for gas associated with partially ionized O. This implies 
that a line-driven stellar wind off
the companion star can not supply a significant fraction of the accreting
material except during the very low luminosity states.  Therefore the
majority of accretion in \lmc\ probably takes place through Roche 
Lobe 
overflow. 

A multi temperature disk blackbody is a good representation
of the soft X-ray spectrum of \lmc\ longward of the disk blackbody peak. Thus
the powerlaw component which dominates at higher energies does not continue
through the soft X-ray range with a powerlaw shape. This suggests that the disk
blackbody component supplies the seed photons which are Compton upscattered to
form the powerlaw at higher energies, as in conventional models for
BHC spectral states.

\section{Acknowledgements}

This work is based on observations obtained with \xmm, an ESA science mission
with instruments and contributions directly funded by ESA Member States and the
USA (NASA). This work makes use of results provided by the ASM/RXTE teams at 
MIT and at the RXTE SOF and GOF at NASA's GSFC.

\end{document}